\documentclass[prd,nofootinbib,preprint,superscriptaddress]{revtex4-1}
\usepackage{amsmath}
\usepackage{graphics}
\usepackage{graphicx}
\pdfoutput=1

\newcommand{\capdef}{}
\newcommand{\mycaption}[2][\capdef]{\renewcommand{\capdef}{#2}%
       \caption[#1]{{\footnotesize #2}}}
\makeatletter
\renewcommand{\fnum@table}{\textbf{\tablename~\thetable}}
\renewcommand{\fnum@figure}{\textbf{\figurename~\thefigure}}
\makeatother

\newcommand {\be}{\begin{equation}}
\newcommand {\ee}{\end{equation}}
\newcommand {\ba}{\begin{eqnarray}}
\newcommand {\ea}{\end{eqnarray}}

\begin{document}


\vspace*{10mm}

\title{Revisiting the quantum decoherence scenario as an explanation for the LSND anomaly\vspace*{1.cm} }

\author{\bf Pouya Bakhti}
\email{pouya_bakhti@ipm.ac.ir} \affiliation{Institute for
	research in fundamental sciences (IPM), PO Box 19395-5531, Tehran,
	Iran}

\author{\bf Yasaman Farzan}
\email{yasaman@theory.ipm.ac.ir} \affiliation{Institute for
research in fundamental sciences (IPM), PO Box 19395-5531, Tehran,
Iran}

\author{\bf Thomas Schwetz}
\email{schwetz@fysik.su.se} \affiliation{Oskar Klein Centre for Cosmoparticle Physics, Department of Physics,
	Stockholm University, SE-10691 Stockholm, Sweden}

\begin{abstract}
  \vspace*{.5cm} We propose an explanation for the LSND anomaly based
  on quantum decoherence, postulating an exponential behavior for the
  decoherence parameters as a function of the neutrino energy.  Within
  this ansatz decoherence effects are suppressed for neutrino energies
  above 200~MeV as well as around and below few MeV, restricting
  deviations from standard three-flavour oscillations only to the LSND
  energy range of 20--50~MeV. The scenario is consistent with the
  global data on neutrino oscillations, alleviates the tension between
  LSND and KARMEN, and predicts a null-result for MiniBooNE. No
  sterile neutrinos are introduced, conflict with cosmology is
  avoided, and no tension between short-baseline appearance and
  disappearance data arises. The proposal can be tested at planned
  reactor experiments with baselines of around 50~km, such as JUNO or
  RENO-50.
\end{abstract}

\maketitle

\section{Introduction}

The three neutrino mass and mixing scheme has been established as the standard paradigm to explain solar, atmospheric, long baseline and reactor experiment neutrino data. There are however some anomalies that cannot be explained within this standard paradigm. Arguably the most significant one is the LSND anomaly \cite{Aguilar:2001ty}. The canonical solution to the LSND anomaly is the four neutrino mixing scheme that includes a sterile neutrino with mass of order of 1~eV and a small mixing with $\nu_e$ and $\nu_\mu$. This solution suffers from some drawbacks. Most importantly, within this scheme there is a tension between appearance and disappearance experiments, see \cite{tension, Giunti:2013aea} for recent analyses. Moreover, a sterile neutrino with mass and mixing parameters to solve the LSND anomaly is in tension with cosmology \cite{Planck:2015xua, cosmol}.

In view of these tensions, quantum decoherence has been suggested in
the literature to explain LSND \cite{Gabriela-04, Gabriela-06,
  Farzan:2008zv}. It is hypothesized that the evolution of quantum
states receives a correction relative to the prediction of standard
quantum mechanics. Such effects could arise in certain quantum
gravity scenarios \cite{Giddings}. To explain the results of short baseline neutrino
experiments, a phenomenological approach is taken to determine
the form and magnitude of decoherence. Neutrino oscillation is a
quantum interference effect over macroscopic distances, which can be
sensitive to small corrections to quantum mechanics. The idea of
involving quantum decoherence to explain the LSND anomaly was first
proposed in \cite{Gabriela-04,Gabriela-06}. As discussed in
\cite{Farzan:2008zv}, within the framework proposed in
\cite{Gabriela-04,Gabriela-06}, the decoherence effects exceed the
upper bound from the NuTeV experiment \cite{NuTeV} in which neutrinos have an
average energy of 75~GeV.

Considering this observation, the so-called
soft decoherence scenario was suggested in \cite{Farzan:2008zv} as a
solution to the LSND anomaly. Within this scenario, the decoherence
effects rapidly decrease with neutrino energy, avoiding the NuTeV
bound and leaving neutrino oscillations in experiments with GeV scale
neutrino energies unaffected. Furthermore, decoherence is restricted
to the 1-3 sector, while the 1-2 sector is not modified. This
explanation does not suffer from the appearance--disappearance tension
of sterile neutrino models. On the other hand, with the power law
energy dependence that was assumed in \cite{Farzan:2008zv}, reactor
neutrinos undergo quantum decoherence after propagating distances of
few cm. This implies that no oscillation due to $\Delta m^2_{31}$ would be possible along the
distance between near and far detectors of reactor experiments. For
this reason, the soft decoherence scenario of \cite{Farzan:2008zv} is
now excluded by the results of the Daya Bay~\cite{An:2013zwz,
  An:2013uza} and RENO~\cite{reno, Kim:2014rfa} experiments.\footnote{Note that the
  DoubleChooz experiment \cite{Abe:2014bwa} does not (yet) exclude
  this scenario, since no data on near--far comparison is available
  to date.} In the present paper, we revisit the decoherence scenario
by modifying the power law assumed in \cite{Farzan:2008zv} to an
exponential energy dependence of the decoherence parameter, leading
to an explanation of LSND consistent with all existing data.

The outline of the paper is as follows. In Sec.~\ref{sec:scenario} we
review the decoherence scenario and introduce the ansatz for the
exponential energy dependence of the decoherence
coefficients. Sec.~\ref{sec:analysis} contains the numerical results
of our analysis of the relevant oscillation data, showing that the scenario can explain
LSND without being in conflict with other data. In
Sec.~\ref{sec:predictions} we discuss further implications of the
scenario and predictions for future experiments. In particular, we
show that planned intermediate baseline (50~km) reactor experiments
can provide a crucial test of the framework. We conclude in
Sec.~\ref{sec:conclusion}.

\section{Quantum decoherence and the LSND anomaly}
\label{sec:scenario}

In the quantum decoherence framework, the evolution of the density matrix for neutrinos can be described as
\begin{equation}
\label{modified-evolution}
\begin{split}
\frac{d\rho}{dt}=-i[H,\rho]-\mathcal{D}[\rho]
\end{split}
\end{equation}
where $H$ is the Hamiltonian and  $\mathcal{D}[\rho]$ parameterizes the decoherence effects. Maintaining complete positivity leads to the Lindblad form for $\mathcal{D}[\rho]$  \cite{Lindblad,Banks}
\begin{equation}
\label{modified}
\begin{split}
{\mathcal{D}}[\rho]=\sum_m \left[ \{ \rho,D_m D_m^\dagger
\}-2D_m\rho D_m^\dagger\right]
\end{split}
\end{equation}
where $D_m$ are  general  complex matrices. Unitarity then requires $D_m$ to be Hermitian. If we further impose conservation of average energy, we find $[H,D_m]=0$. As a result, in the neutrino mass basis, $D_m$ and $H$  can be simultaneously diagonalized
\begin{equation}
\label{form}
\begin{split}
H={\rm Diag}[h_1,h_2,h_3] \,,
\qquad
D_m={\rm	Diag}[d_{m,1},d_{m,2},d_{m,3}]\,,
\end{split}
\end{equation}
where $h_i=(p^2+m_i^2)^{1/2}$ (adopting the equal momentum approximation for the mass states), and $d_{m,i}$ are unknown energy dependent real quantities with dimension of $[mass]^{1/2}$. Solving Eq.~(\ref{modified-evolution}), we find
\begin{equation}
\label{evolved}
\begin{split}
\rho(t)=
\left[ \begin{matrix}
\rho_{11}(0)&\rho_{12}(0) e^{-(\gamma_{12}-i\Delta_{12})t}&
\rho_{13}(0)e^{-(\gamma_{13}-i\Delta_{13})t}
\cr \rho_{21}(0) e^{-(\gamma_{21}-i\Delta_{21})t}&
\rho_{22}(0)&\rho_{23}(0) e^{-(\gamma_{23}-i\Delta_{23})t}\cr
\rho_{31}(0) e^{-(\gamma_{31}-i\Delta_{31})t}&\rho_{32}(0)
e^{-(\gamma_{32}-i\Delta_{32})t}&\rho_{33}(0)
\end{matrix} \right]
\end{split}
\end{equation}
in which
\begin{equation}
\label{gd}
\begin{split}
\gamma_{ij} \equiv \sum_m(d_{m,i}-d_{m,j})^2 \ \ {\rm and} \ \
\Delta_{ji} \equiv  h_j-h_i \approx \frac{\Delta m_{ji}^2}{2E_\nu} \,.
\end{split}
\end{equation}
Obviously, $\gamma_{ij} = \gamma_{ji}$ and
$\Delta_{ij} = -\Delta_{ji}$.
This means $\gamma_{ij}$ is symmetric under flipping $i \leftrightarrow j$. In the following,
we assume that only one term contributes in the sum and we drop the index $m$.
The flavor conversion probability can be written as
\begin{equation}
\label{eq:prob}
\begin{split}
P_{\alpha \beta} = \langle \nu_\beta |
\rho^{(\alpha)}(t) |\nu_\beta\rangle =  \sum_{ij} U_{\beta i}^*
U_{\beta j} \, \rho_{ij}^{(\alpha)}(t) \,
\end{split}
\end{equation}
where $U_{\alpha i}$ are the elements of the PMNS matrix \cite{Maki:1962mu}. The density matrix $\rho_{ij}^{(\alpha)}(t)$ is given by Eq.~(\ref{evolved}) and  $\rho_{ij}(0) = \rho^{(\alpha)}_{ij}(0) = U_{\alpha i} U_{\alpha j}^*$. The flavor conversion probability for antineutrinos, $P_{\bar{\alpha}\bar{\beta}}$, will be given by a similar formula, replacing $U$ with $U^*$.

In the soft decoherence scenario of \cite{Farzan:2008zv}, a power law energy dependence of the decoherence coefficients has been assumed, $d_i \propto E^{-r}$ ($r \ge 2$), suppressing decoherence effects for $E \gtrsim 100$~MeV. However, as mentioned in the introduction, this leads to strong decoherence effects at low energies and is by now excluded by Daya Bay and RENO results. In this work, we therefore propose a modified energy dependence of the decoherence parameters and we conjecture an exponential dependence on energy for $d_{i}$ as follows:
\begin{equation}
\label{di}
\begin{split}
d_{i}=\sqrt{\gamma_{0}} \ \exp\left[{-\left(\dfrac{E}{E_i}\right)^{n}}\right] \,,
\end{split}
\end{equation}
where $\gamma_{0}$ is a constant parameter with dimension of mass, universal for all mass eigenstates. $E_{i}$ are also constant parameters with dimension of mass but can in principle take different values for different mass eigenstates. The power  $n$ can take any arbitrary number.
In line with the idea of soft decoherence, we take a value for $n$ and $E_i$ for which at energies $\gtrsim {\rm few} \times 100$~MeV, the decoherence parameters become suppressed rapidly enough not to have any effects at experiments such as MINOS~\cite{Adamson:2014vgd}, T2K~\cite{Abe:2013hdq}, atmospheric neutrinos~\cite{Ashie:2004mr} and etc. In the same way this predicts null-results for short-baseline experiments with $E \gtrsim 200$~MeV such as MiniBooNE~\cite{AguilarArevalo:2007it, Mahn:2011ea} , CDHS~\cite{Dydak:1983zq}, NOMAD~\cite{NOMAD}, NuTeV~\cite{NuTeV} and etc. We found that with $E_i <100$ MeV and $n\geq 2$, this requirement is fulfilled. Unless it is stated otherwise, we take $n=2$ for definiteness throughout this paper.

To avoid constraints from the long-baseline KamLAND reactor experiment  \cite{kamland}, we restrict the scenario to $d_1\approx d_2$ or equivalently to $\gamma_{12} \approx 0$ \cite{Farzan:2008zv}. In the limit $|E_1 - E_2| \ll E_1$ with taking $n=2$ and $E \lesssim E_1$, we find $\gamma_{12}\simeq 4 \gamma_0 \exp(-2E^2/E_1^2) E^4 (E_2-E_1)^2/E_1^6$. To avoid bounds from KamLAND, $\gamma_{12}$ should be much smaller than $\sim (200~{\rm km})^{-1}$ at $E\sim$ few MeV which for $E_1 \gg $~MeV means $|E_2-E_1|/E_1 \ll (800~{\rm km} ~\gamma_0)^{-1/2} [E_1/({\rm few ~MeV})]^2$.  At first sight, it seems that from solar neutrino data, we can obtain strong bounds on $\gamma_{12}$, too. However, for long baselines, the interference effects are averaged out and as a result the sensitivity to $\gamma_{ij}$ is lost. This happens for solar neutrinos even before reaching the resonance region inside the Sun. From a theoretical point of view, it may be natural to assume that $d_i$ are functions of mass: $d_i=f(m_i)$. From $m_1\simeq m_2 \ne m_3$, we then expect $d_1\simeq d_2 \ne d_3$.

In the rest of this paper, we shall take
\begin{equation}
\label{gam}
\begin{split}
\gamma_{12}=0 \ \ {\rm and} \ \
\gamma\equiv\gamma_{13}=\gamma_{32}=\gamma_{0}\left(\exp\left[{-\left(\frac{E}{E_3}\right)^{n}}\right]-\exp\left[{-\left(\dfrac{E}{E_1}\right)^{n}}\right]\right)^2\ ,
\end{split}
\end{equation}
with $n=2$.
Notice that the combination in the parenthesis is less than or equal to $1$ and hence,
$\gamma \leq \gamma_0$. For $E\gg E_1,E_3$, we have $d_1,d_3 \to 0$ and $\gamma$ will therefore exponentially converge to zero. For $E\ll E_1,E_3$, $\gamma$ will also be small and suppressed by $[E^n(E_1^{-n}-E_3^{-n})]^2$. Only for $E\sim E_1,E_3$, the value of $\gamma$ can be sizable and decoherence effects can be significant. Note that the suppression of decoherence at low energies works only for a universal coefficient $\gamma_0$. Hence, the assumption that $\gamma_0$ is independent of the neutrino mass is crucial for our scenario.

For  $\Delta_{21}L \ll 1$, we can write
\begin{eqnarray} \label{pr}  P_{\bar{\mu} \bar{ e}}(\gamma,L) =
P_{\mu e}(\gamma,L) &=& P_{e \mu}(\gamma,L)\simeq 2|U_{\mu 3}|^2 |U_{e
	3}|^2\left[1- e^{-\gamma L} \cos (\Delta_{31}L) \right]\, \\  P_{\bar{e}\bar{e}}(\gamma,L) =
P_{ee}(\gamma,L) &\simeq & 1-2|U_{e3}|^2(1-|U_{e3}|^2)\left[1- e^{-\gamma L}
\cos (\Delta_{31}L) \right]\,,
\label{probabilities} \\ P_{\bar{\mu} \bar{\mu}}( \gamma,L) =
P_{\mu \mu}( \gamma,L) &\simeq & 1-2|U_{\mu 3}|^2(1-|U_{\mu 3}|^2)\left[1-
e^{-\gamma L} \cos (\Delta_{31}L) \right] \,.
\end{eqnarray}
For $\gamma L \to 0$ the quantum decoherence is turned off and the  flavor conversion probability becomes equal to that in the  standard three neutrino oscillation scenario.

\section{Analysis of short baseline and reactor neutrino data}
\label{sec:analysis}

In this section, we present the results from a numerical analysis of
relevant data and determine the allowed range of parameters which can
account for the LSND anomaly without being in conflict with any other
experimental results.

\subsection{Description of the used data and analysis details}

In our analysis, we focus on the LSND electron antineutrino excess
events in the energy range from 20~MeV to 60~MeV
\cite{Aguilar:2001ty}. We extract the data points as well as the
background from Fig.~24 of \cite{Aguilar:2001ty}. The data sample
shown in that figure was obtained by applying the analysis cut
$R_\gamma> 10$, see section VII-C of \cite{Aguilar:2001ty} for the
definition of the $R_\gamma$ variable. To predict the number of events
in each bin within the decoherence scenario, we normalize the total
number of events for $P(\bar{\nu}_\mu \to \bar{\nu}_e)=1$ to 33300 as
indicated in table~VIII of \cite{Aguilar:2001ty}, multiplied by 0.39
which is the efficiency of the $R_\gamma>10$ cut (see table~IX of
\cite{Aguilar:2001ty}). The $\chi^2$ is defined as the sum of squares
of the difference between prediction (signal+background) and observed
number of events per bin divided by the square of the uncertainty. The
sum is over bins with $20~{\rm MeV}<E<60$~MeV (10 bins) and the
uncertainties in each bin are obtained from the error bars on the data
points in Fig.~24 of \cite{Aguilar:2001ty}, which account for both
systematic and statistical uncertainties. We have checked that our
analysis reproduces the allowed region for standard oscillations
obtained in \cite{Aguilar:2001ty} with good accuracy.

We also take into account the results of the KARMEN experiment
\cite{karmen}, which observes 15 events in the energy range from
16~MeV to 52~MeV with a predicted background of $15.8\pm0.5$
events. Any explanation of the LSND anomaly has to address the
null-result of KARMEN, taking into account the very similar
experimental configuration, with the main difference being the
somewhat shorter baseline of KARMEN. Again we perform a fit to the
binned energy spectrum (9 bins) and we can reproduce the official
results in terms of sterile neutrino oscillations to good accuracy.
For short baseline experiments such as LSND and KARMEN, $\Delta_{31}L
\ll 1$, so we can  use Eq.~(\ref{pr}) to  write the conversion
probability for neutrinos and antineutrinos as follows:
\be
P_{\bar{\mu} \bar{e}}(\gamma,L) = P_{\mu e}(\gamma,L) = 2|U_{\mu 3}|^2
|U_{e 3}|^2 \left(1- e^{-\gamma L} \right) \approx |U_{e 3}|^2\left(1-
e^{-\gamma L}\right) \,.
\ee

As discussed before, the data from Daya Bay and RENO, being consistent
with the standard three neutrino oscillation scheme, can put bounds on
the decoherence parameters $E_1$ and $E_3$. To derive the bounds, we
analyze the energy spectrum of the $\bar{\nu}_e$ flux at the near and
far detectors of Daya Bay shown in Fig.~2 of \cite{An:2013zwz}. We
read the data points for near detectors (EH1 and EH2) and far detector
(EH3) from the upper panel in pairs of panels shown in Fig.~2 of
\cite{An:2013zwz}, 75 data points in total. We read the background for each detector from the
inset panels in this figure.  Finally, having extracted the data and
background, to calculate the number of events per bin without
oscillation ({\it i.e.,} for $P(\bar{\nu}_e\to \bar{\nu}_e)=1$), we
use the data points displayed in the lower panels
[(data--background)/predictions] of Fig.~2 of \cite{An:2013zwz}.
To calculate the number of events within the decoherence scenario, we
then multiply this number with the probability in
Eq.~(\ref{probabilities}), averaged over cross section, flux, and energy resolution.
To compute the $\chi^2$, we equate uncertainties
for each bin to the root of number of events per bin ({\it i.e.,} the
statistical uncertainty). The overall flux normalization is taken to be a free
parameter to be fixed by the combined near and far detector fit.
The distances between the various reactor and
detector sites of the Daya Bay experiment are taken from table~2 of
\cite{An:2013uza}.

We also include the spectrum of fifty thousand inverse beta-decay
candidate events of the
far detector of the RENO experiment \cite{reno} and compare it with
the prediction. Data points are taken from the right panel of Fig.~4
of \cite{Kim:2014rfa} (26 data points), where the background is already subtracted. To
compute the prediction of the decoherence scenario for each bin we
multiply the oscillation prediction shown in Fig.~4 of
\cite{Kim:2014rfa} by the averaged survival probability in
Eq.~(\ref{probabilities}) and divide by the oscillation probability with
$\sin^2 2 \theta_{13}=0.094$ and $|\Delta m_{31}^2|=2.32\times
10^{-3}~e{\rm V}^2$ as stated in Fig.~4 of \cite{Kim:2014rfa}.  These
mass and mixing parameters are the best fit values that
Ref.~\cite{Kim:2014rfa} derives by using a MC simulation to fit both
near and far detector data. For values of $E_1$ and $E_3$ of interest
for solving the LSND anomaly ($E_1\sim E_3\sim {\rm few}~ 10$~MeV),
decoherence at the near detector is negligible ($\gamma L \ll 1$) and
Eq.~(\ref{probabilities}) converges to the standard oscillation
formula. Hence, using the far detector prediction based on the
near detector data (as done for Fig.~4 of \cite{Kim:2014rfa}) should
be a good approximation. Notice, however, that including the RENO
results does not much change the overall results for the decoherence
fit, which is dominated by Daya Bay data.

\subsection{Results of our fit}

\begin{figure}[t] \centering
	\includegraphics[width=0.48\textwidth]{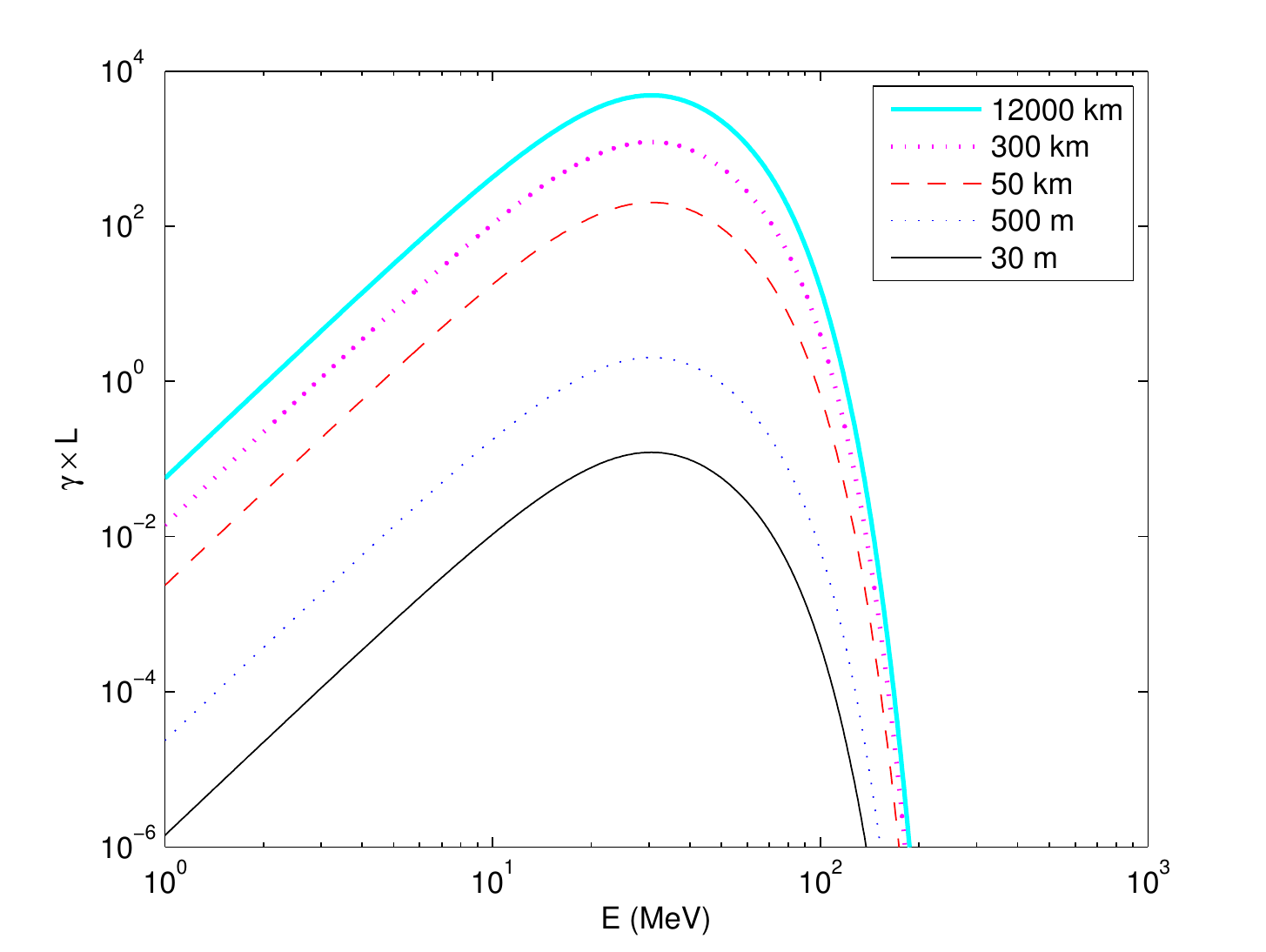}
        \includegraphics[width=0.48\textwidth]{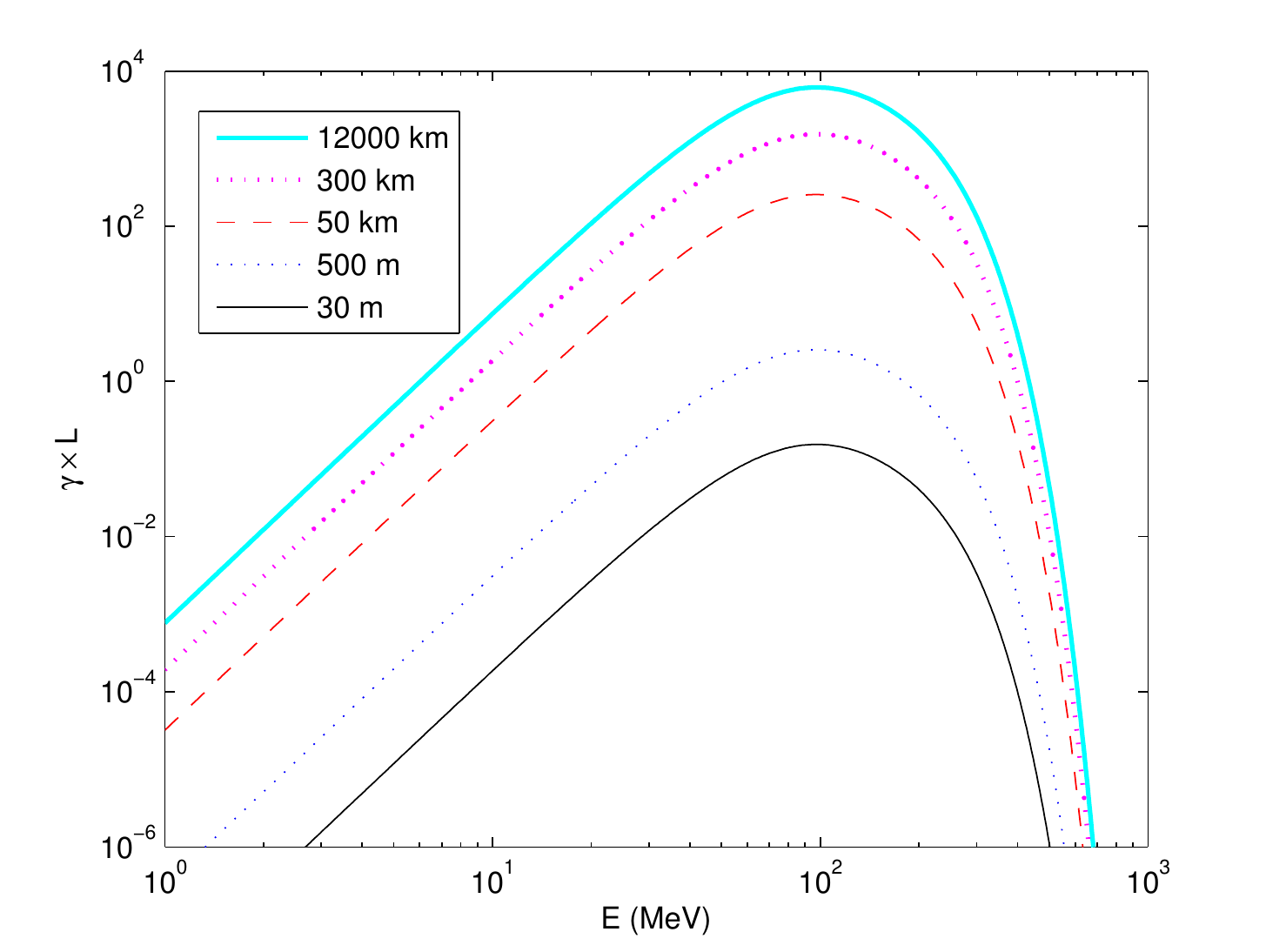}
        \mycaption{\label{gle} Dependence of $\gamma L$ on energy for
          different baselines corresponding approximately to LSND,
          MiniBooNE, medium baseline reactor, long baseline
          accelerator and atmospheric neutrino experiments. We have
          taken $n=2$, $\gamma_0=0.01$~m$^{-1}$ for both panels, and
          $E_1=E_2=20$~MeV, $E_3=55$~MeV ($E_1=E_2=60$~MeV,
          $E_3=200$~MeV) for the left (right) panel.}
\end{figure}

Remember that $P_{\bar{\mu}\bar{e}}$ at LSND should be of order
few$\times 10^{-3}$ to account for the observed excess. For
$|U_{e3}|^2 \simeq 0.02$, the value of $\gamma L$ for LSND should be
of order of 0.1 to explain the anomaly. From $\gamma<\gamma_0$ and
$L=30$~m, we find that $\gamma_0$ has to be of order of 0.01~m$^{-1}$
or larger. Larger values of $\gamma_0$ require $E_1 \simeq E_3$ to
cause partial cancelation, see Eq.~(\ref{gam}). To explain the LSND
anomaly, we demand that $\gamma \sim \gamma_0\sim 0.01$~m$^{-1}$ at
$E\sim 30$~MeV and to avoid the bounds from reactor experiment as well
as from higher energy experiments, we require $\gamma \ll \gamma_0$
for both $E\gg 30$~MeV and $E\sim$~few~MeV. That means $E_1$ and $E_3$
should be of order of 10~MeV. Fig.~\ref{gle} shows $\gamma L$ versus
energy taking typical values for decoherence parameters. The left
panel of that figure corresponds to a parameter choice close to the
best fit value of our model.  As seen from Fig.~\ref{gle} (left), at
$\gamma_0 =0.01$~m$^{-1}$, the effect of decoherence is negligible for
energies above 200~MeV. Thus, the bounds from short-baseline
experiments such as NOMAD, CDHS, or NuTeV are satisfied.  In other
words, like in the soft decoherence scenario, the tension between
appearance and disappearance experiments plaguing the 3+1 sterile
oscillations is solved. Furthermore, the standard oscillation results
for experiments with $\mathcal{O}$(1~GeV) neutrinos such as MINOS,
T2K, or atmospheric neutrinos are not affected. For $E_i > 200$~MeV
(see right panel of Fig.~\ref{gle}), decoherence effects can
potentially show up in the low energy bins of T2K as well as in the
sub-GeV atmospheric neutrino data.

\begin{figure}[t] \centering
	\includegraphics[width=0.6\textwidth]{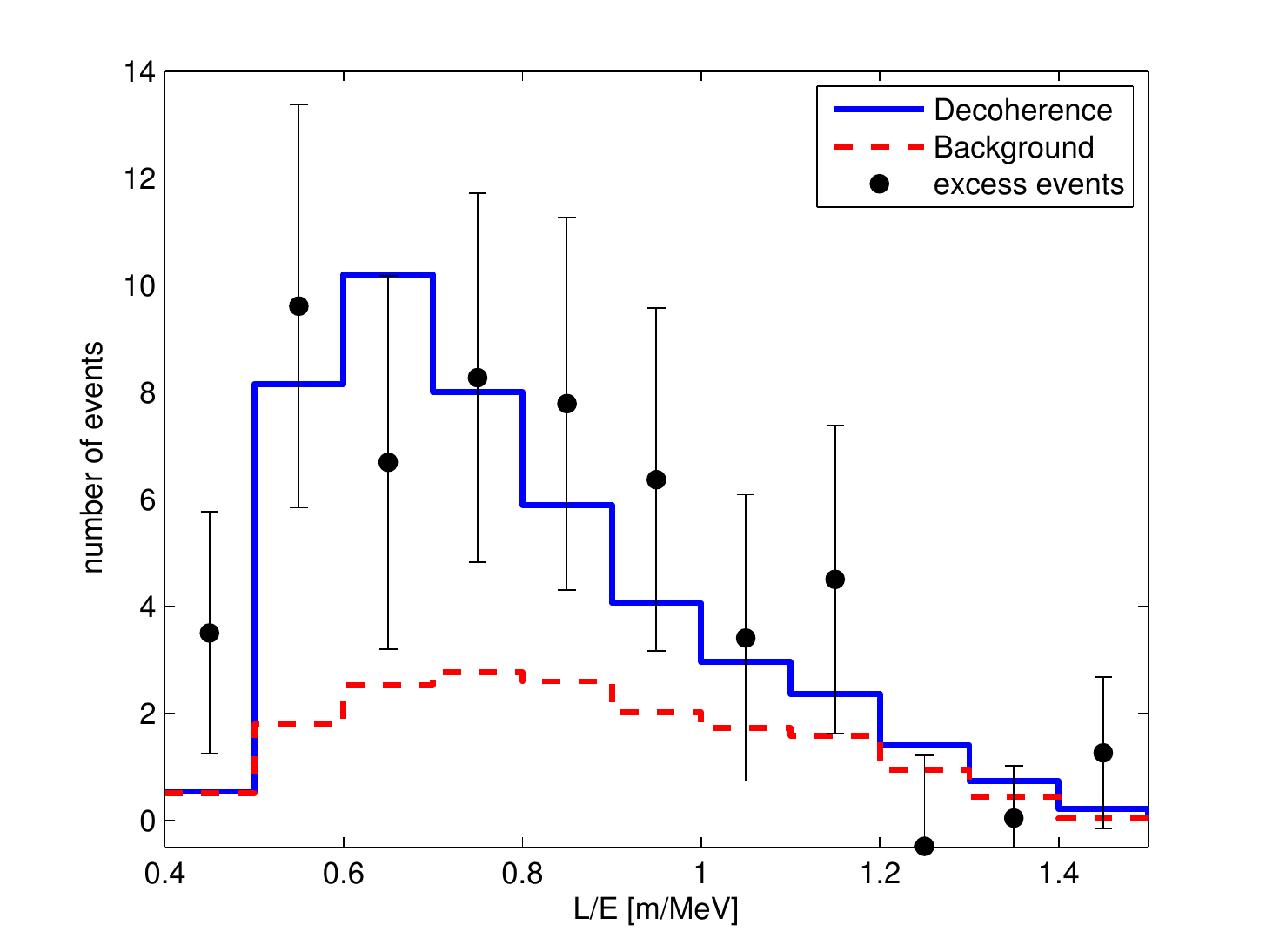}
        \mycaption{\label{LSND} Decoherence prediction for LSND for
          $\gamma_0=0.01$~m$^{-1}$, $E_1=E_2=18$~MeV and $E_3=63$~MeV
          compared with data.    }
\end{figure}

Our main focus is on a range of parameters for which reactor and T2K experiments are unaffected. As a result, a reanalysis of Daya Bay and T2K will approximately yield the same value for $\theta_{13}$ as in the standard oscillation case. We fix the values of the standard neutrino parameters (including $\theta_{13}$) to the best fit value of the global analysis from \cite{Gonzalez-Garcia:2014bfa}. We find that within this scenario with $n=2$, LSND data can be explained with a satisfactory p-value of 68~\% with three unknown parameters fitted to $\gamma_0=0.01$~m$^{-1}$, $E_1=E_2=18$~MeV and $E_3=63$~MeV.  {The spectrum of events at LSND for these values is shown in Fig.~\ref{LSND}. The figure demonstrates  that data and prediction of the decoherence scenario are in good agreement.} In the following analysis we will fix $\gamma_0$ to the LSND best fit value of 0.01~m$^{-1}$.

\begin{figure}[t] \centering
	\includegraphics[width=0.6\textwidth]{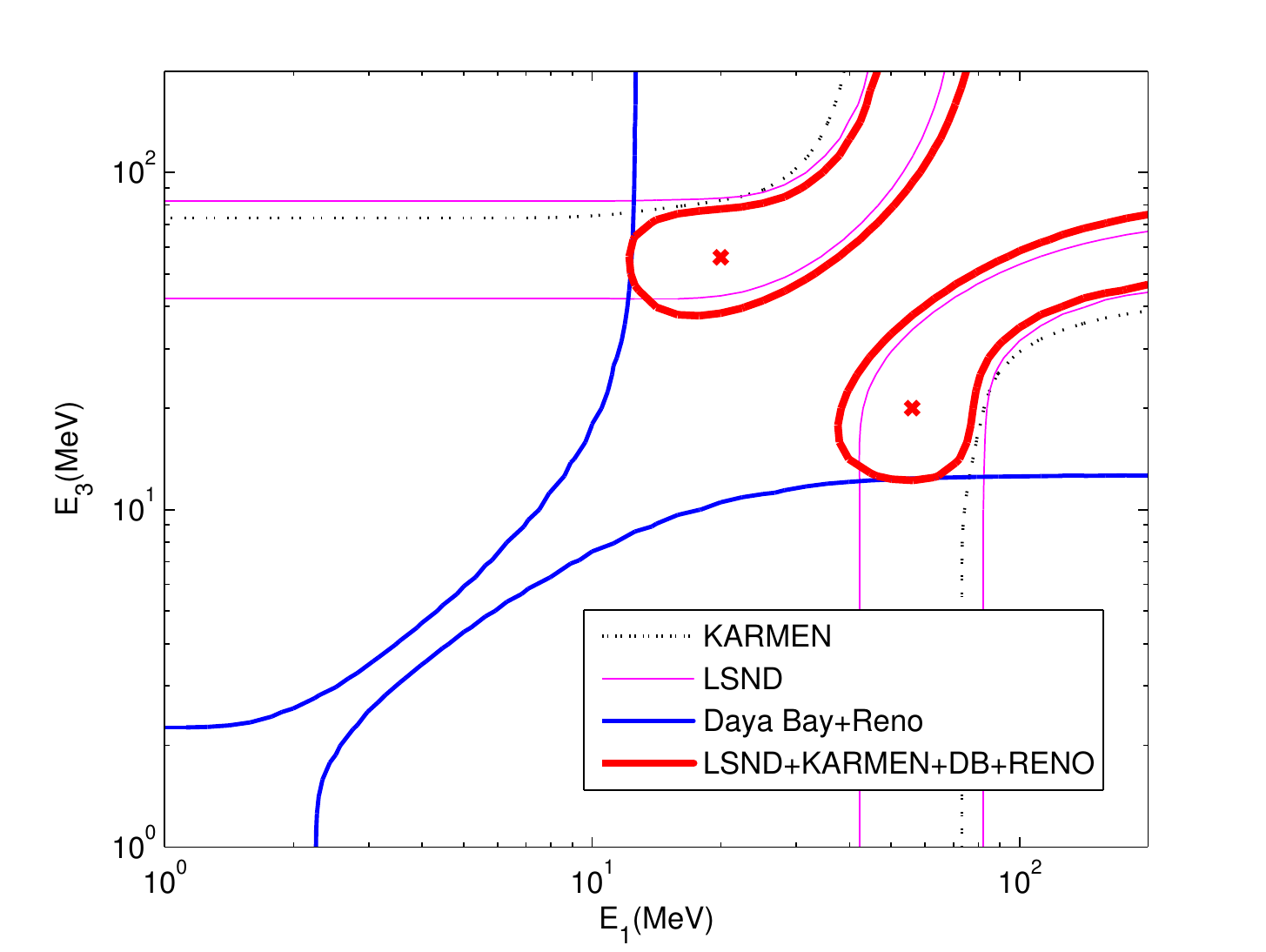}
        \mycaption{\label{Ei} Constrains on the parameters $E_{1,3}$
          from short baseline and reactor experiments at
          90\%~C.L.\ taking $n=2$ and $\gamma_0=0.01$~m$^{-1}$. The
          region below and to the left of the dotted curves is allowed
          by KARMEN, the region between the dark-blue solid curves is
          allowed by Daya Bay and RENO, the thin magenta curves
          delimit the regions allowed by LSND. The regions
          consistent with all data are inside the thick red curves,
          with the cross indicating the best fit point.}
\end{figure}

Fig.~\ref{Ei} shows the constraints from short baseline and reactor
neutrino experiments on $E_1$ and $E_3$ at 90~\%~C.L., fixing $n=2$
and $\gamma_0=0.01$~m$^{-1}$. As expected the bounds are symmetric
under $E_1 \leftrightarrow E_3$. Due to the exponential dependence on
the distance $L$, the difference in the baselines for LSND and KARMEN
(30~m versus 18~m) leads to a better consistency of the two results
than in the case of oscillations. Fig.~\ref{Ei} shows that at
90\%~C.L.\ KARMEN only marginally constrains the LSND allowed region,
compare thin magenta (LSND) and dotted black (KARMEN) curves. For
$E_i\lesssim 8$~MeV, the bound from the reactor neutrino experiments
becomes stringent and practically only the narrow region with
$E_1\simeq E_3$ is allowed. But for $E_1,E_3>15$~MeV, the bounds from
reactor experiments are relaxed. The thick red curves in the plot show
the globally allowed region. The best fit is marked by a cross in the
plot and it is located at $E_1 = 20$~MeV and $E_3 = 56$~MeV. We
have clipped the figure at $E_1,E_3=200$~MeV because for larger values
T2K and atmospheric neutrinos will also be affected, as visible from
the right panel of Fig.~\ref{gle}.

\begin{table} [t] \centering	
\begin{tabular}{l*{6}{c}r}
        \hline\hline
	Data               &$ \chi^2_{\rm min}/$DOF & GOF & $\chi ^2_{\rm PG}/$DOF & PG  \\
	\hline
	LSND & 4.8/8 & 77\%   \\
	KARMEN  &  7.0/7 & 43\%          \\
	Daya Bay and RENO & 78/98 & 93\%           \\
	LSND+KARMEN     & 14/17 & 66\% & 2.3/2 & 32\% \\
	LSND+KARMEN+Reactor  & 93/118 & 96\% & 3.2/4 & 52\% \\
\hline\hline
\end{tabular}
  \mycaption{\label{tab:PG} $\chi ^2_{\rm min}/$DOF and goodness of
    fit (GOF) for different combinations of short baseline and reactor
    neutrino data. The last two columns quantify the consistency of
    different experiments, see Eq.~(\ref{eq:PG}) for the
    definition. $E_1=E_2$ and $E_3$ are taken as free parameters to
    fit the data and the rest are fixed to $\gamma_0=0.01$~m$^{-1}$,
    $n=2$ and $\sin^2 2 \theta_{13}=0.085$.}
\end{table}

Table \ref{tab:PG} shows $\chi^2_{\rm min}$ per degrees of freedom and
goodness of fit (GOF or p-value) for various short baseline and
reactor neutrino experiments. Notice that the p-value for LSND given
in the table ({\it i.e.,} for the case that $\gamma_0$ is fixed and
$E_1$ and $E_3$ are treated as free parameters) is better than the
aforementioned p-value that we obtain when we treat $\gamma_0$ as a
free parameter along with $E_1$ and $E_3$. This reflects the fact that
for $\gamma_0>0.01$~m$^{-1}$, the minimum value of $\chi^2$ over the
$E_1$ and $E_3$ plane does not change much by varying the value of
$\gamma_0$. Let us comment on the somewhat large p-value of 93\% for
the reactor analysis. If Daya Bay and RENO are analyzed separately we
find $\chi^2_{\rm min}$/DOF values of 55/72 and 23/24,
respectively. Hence, the too good fit comes from the Daya Bay
analysis. This might be related to the accuracy of reading data from
the plot. Note however, that our results are based on
$\Delta\chi^2$ values, which are insensitive to the absolute value of
the $\chi^2$. Furthermore, we can reproduce the standard $\theta_{13}$
result of Daya Bay with good accuracy.

Consistency of the combination of various  experiments is quantified  by the so-called Parameter Goodness of fit (PG)~\cite{Maltoni:2002xd, Maltoni:2003cu} defined as
\begin{equation} \label{eq:PG}
\chi^2_\text{PG} =
\chi^2_\text{tot,min} - \sum_i \chi^2_{i,\text{min}} \,,
\end{equation}
where $\chi^2_\text{tot,min}$ is the global minimum, the sum over $i$ runs over the different experiments, and $\chi^2_{i,\text{min}}$ are the minima of the experiments separately.
As seen from the third and fourth columns of table~\ref{tab:PG}, the KARMEN and LSND data are in good agreement with each other under the decoherence hypothesis, thanks to the exponential dependence of the transition probability on $L$. Furthermore,  the short-baseline experiments LSND and KARMEN are also in very good agreement with the Daya Bay and RENO reactor experiments.

From Fig.~\ref{gle}, it is clear that decoherence effects are strongly
suppressed for the MiniBooNE baseline of around 500~m and neutrino
energies above 200~MeV. Thus, our scenario is consistent with a
null-result in MiniBooNE, as observed in the energy range $E >
475$~MeV~\cite{AguilarArevalo:2007it}. The low-energy event excess
between 200 and 475~MeV~\cite{AguilarArevalo:2008rc} is not explained
since the transition probability is already highly suppressed in that
regime. It is necessary to be consistent with T2K and atmospheric
neutrino data, which requires us to restrict $E_{1,3}$ to values
sufficiently low such that  not to affect the standard oscillation behavior
seen there. Hence the low-energy MiniBooNE excess has to find an
alternative explanation.

\section{Predictions for future experiments and possible experimental tests}
\label{sec:predictions}

First we mention that the so-called reactor \cite{rec-an} and Gallium
\cite{Giunti:2010zu} anomalies cannot be explained in the decoherence
framework proposed here. At reactor energies and below, the decoherence
effects are suppressed so we predict neither a reduced reactor
neutrino flux at short baselines, nor a reduced neutrino rate in
source experiments at Gallium detectors. Those anomalies (which are at
the level of $3\sigma$) should find another explanation in the
scenario discussed here. Planned experiments at reactors with very
short baselines as well as radioactive source experiments should lead
to null-results.

Let us now comment on future accelerator-based experiments.
Long-baseline oscillation experiments such as
NOvA~\cite{Patterson:2012zs} or LBNF~\cite{lbnf}  use neutrino
beams with $E_\nu \gtrsim 1$~GeV. As clear from Fig.~\ref{gle} we
predict no decoherence effects at those energies and hence such
experiments should obtain results consistent with standard
three-flavour oscillations. The nuSTORM short baseline neutrino
experiment with an average energy of 3~GeV and a baseline of 2~km is
proposed to test the 3+1 oscillation hypothesis
\cite{Adey:2014rfv}. From Fig.~\ref{gle}, we observe again that the
decoherence effects for this setup are too small so we predict a null
signal for such an experiment. If nuSTORM finds no signal for
appearance, the 3+1 solution will be ruled out but the decoherence
solution will still survive, while the observation of an appearance
signal at nuSTORM would exclude the decoherence solution proposed
here. The situation is similar also for other short baseline neutrino
experiment proposals with neutrino energies $\gtrsim {\rm few} \times
100$~MeV, see e.g., \cite{sbl-future}. For the ESS superbeam
\cite{ess}, with a peak energy of $E\simeq 200$~MeV some
decoherence effects may start to show up if the $E_{1,3}$ parameters
are not too small. We do predict an appearance signal in LSND-like
short baseline experiments with energies around 30~MeV; see e.g.,
\cite{Elnimr:2013wfa}.

A crucial test of our scenario might be possible with reactor
experiments with baselines of around 50~km, such as the JUNO
\cite{He:2014zwa, Li:2013zyd} and RENO-50 \cite{Kim:2014rfa} projects. Using the
formalism of Sec.~\ref{sec:scenario} we obtain for the three-flavour
$\bar\nu_e$ survival probability
\begin{equation}\label{PeeJUNO}
  \begin{split}
P_{\bar e \bar e} &= 1 - \sin^22\theta_{12}\sin^2\frac{\Delta_{21} L}{2}
- \frac{1}{2}\sin^22\theta_{13} \\
&+
\frac{1}{2}\sin^22\theta_{13} \, e^{-\gamma L}
\left[ \cos^2\theta_{12} \cos(\Delta_{31}L) + \sin^2\theta_{12} \cos(\Delta_{32}L)\right]\,.
  \end{split}
\end{equation}
The main goal of those experiments is to observe the small
``wiggles'' in the energy spectrum induced by the term in the second
line of Eq.~\eqref{PeeJUNO}. We see that for $\gamma L \gtrsim 1$
those features will be suppressed due to decoherence.  Fig.~\ref{gle}
(left) shows that for baselines of 50~km, $\gamma L$ becomes of order
one for $E_\nu \gtrsim 4$~MeV, and hence, the fast oscillations in the
survival probability may become suppressed.

We perform a numerical study of this effect by simulating the JUNO
experiment, using information from \cite{Li:2013zyd}.  We normalize
the number of events such that for the default exposure of 20~kt
detector mass $\times$ 36~GW reactor power $\times$ 6~yr exposure
(4320~kt~GW~yr in total) we obtain $10^5$ events. The energy
resolution is assumed to be $3\% \sqrt{1\,{\rm MeV}/E_\nu}$. We
perform a $\chi^2$ analysis using 350 bins for the energy
spectrum. Several systematic uncertainties are included as well as the
smearing induced by the baseline distribution of 12 relevant reactor
cores. Further details of our analysis can be found in
\cite{Blennow:2013oma}.

\begin{figure}[t] \centering
	\includegraphics[height=0.35\textwidth]{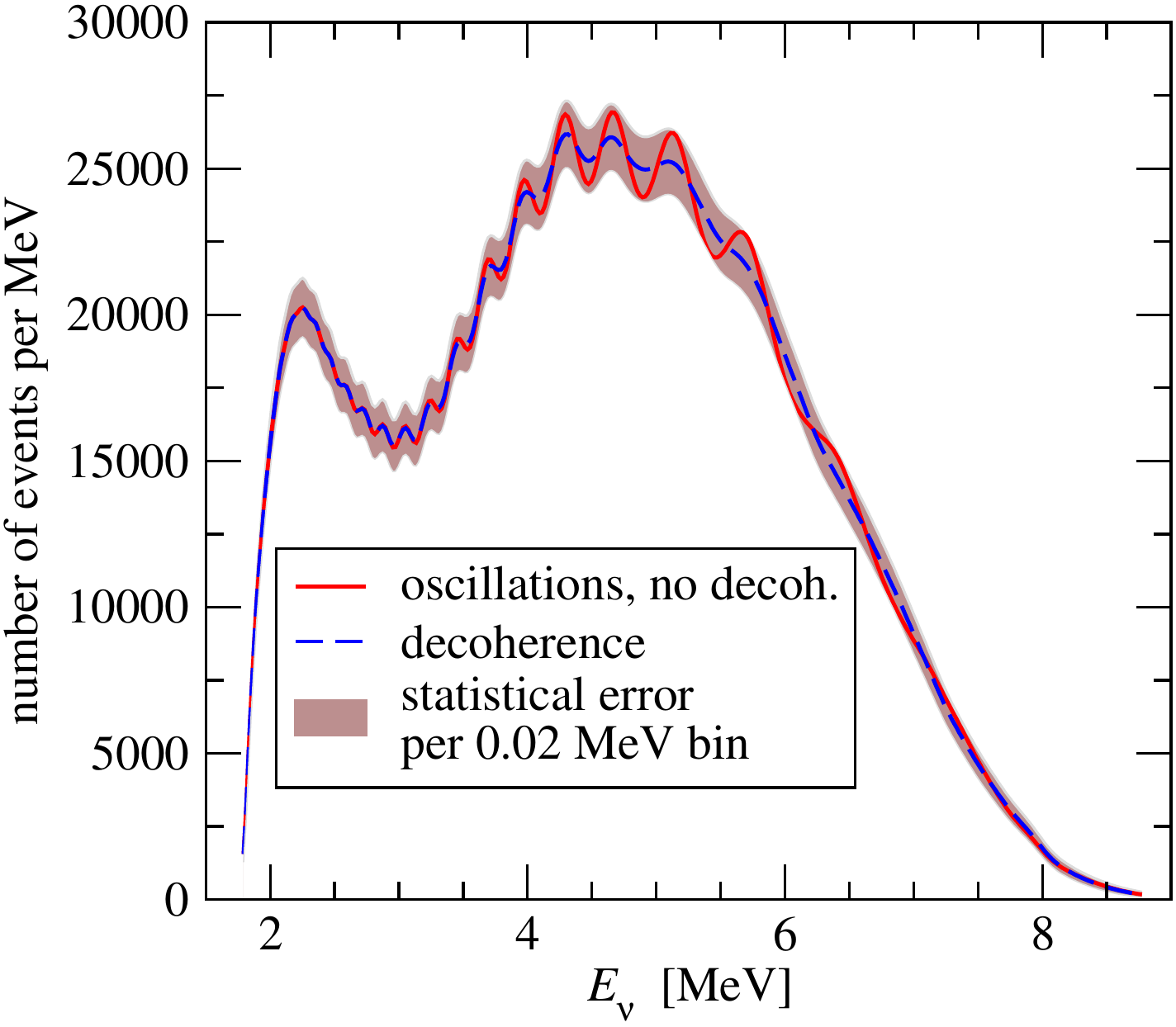}
        \includegraphics[height=0.35\textwidth]{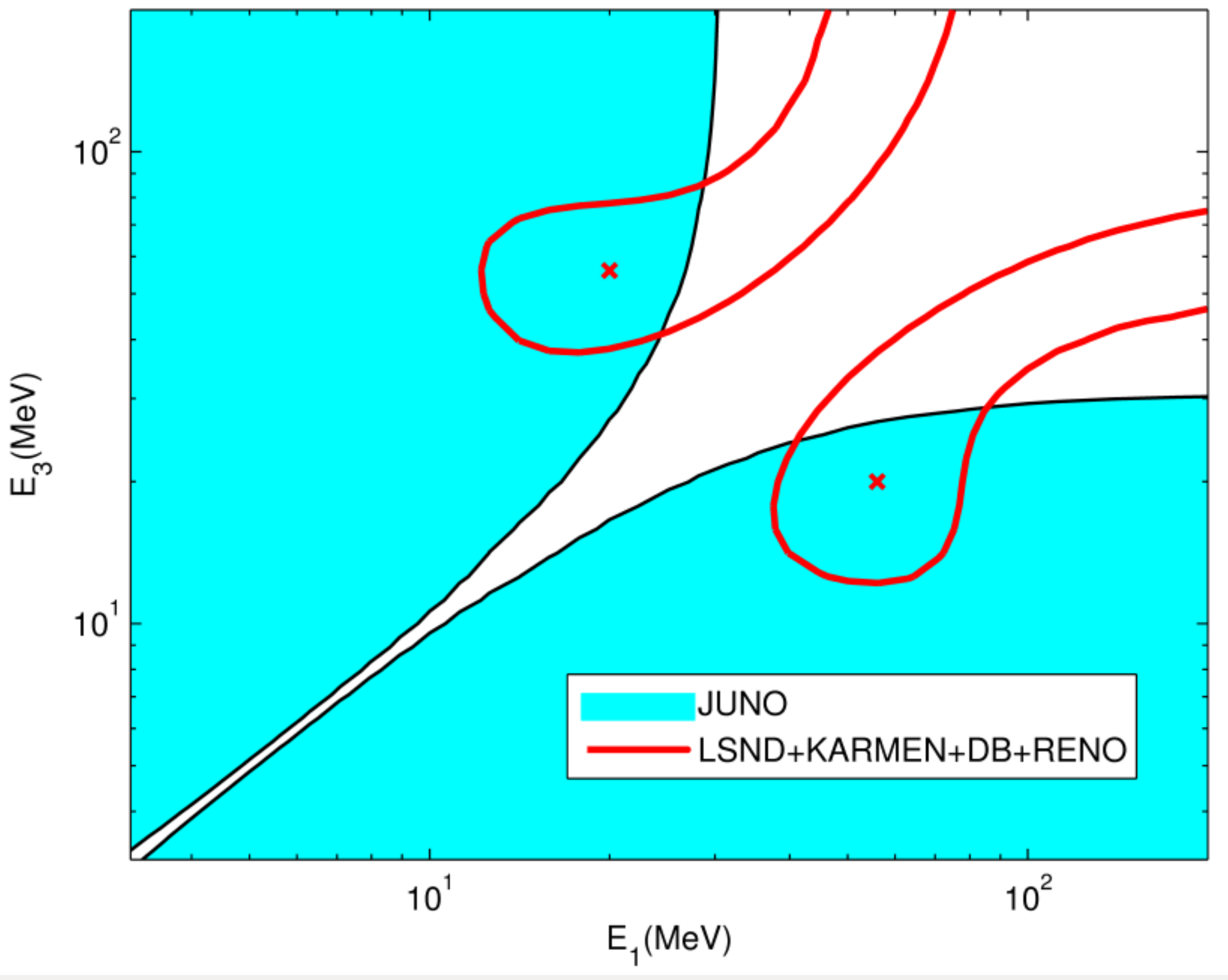}
        \mycaption{\label{juno} Left: Event spectrum at JUNO for an
          exposure of 4320~kt~GW~yr. The red solid curve corresponds
          to standard oscillations with $\Delta m_{31}^2>0$, whereas the blue dashed curve
          shows the spectrum with decoherence parameters $E_1=E_2 =
          20$~MeV, $E_3 = 55$~MeV, $\gamma_0=0.01$~m$^{-1}$,
          $n=2$. The shaded band indicates the statistical error per
          0.02~MeV bin. Right: In the shaded regions, JUNO can distinguish the decoherence
          scenario from standard oscillations at more than $3\sigma$
          ($\Delta\chi^2 = 9$). The red curves show the 90\%~C.L.\ allowed regions
          from the combined analysis of LSND, KARMEN, Daya Bay and RENO,
          with the cross indicating the best fit point (same as in
          Fig.~\ref{Ei}).}
\end{figure}

In Fig.~\ref{juno} (left) we compare the expected spectra for standard
oscillations (red solid) to the decoherence scenario with parameters
close to the best fit point (blue dashed). We clearly observe that the
$\theta_{13}$-induced modulation of the spectrum (second line in
Eq.~\eqref{PeeJUNO}) becomes suppressed in the case of decoherence for
neutrino energies above 4~MeV. Thanks to the huge number of events
this difference is highly significant: the $\Delta\chi^2$ between
those two curves is 33, which means that the no-decoherence hypothesis
would be excluded at more than $5\sigma$ (1~DOF, i.e., for fixed
parameters). The right panel of Fig.~\ref{juno} shows the regions in
the ($E_1, E_3$) plane, where JUNO will be able to distinguish the
decoherence scenario from standard oscillations at more than $3\sigma$
($\Delta\chi^2 = 9$). We observe that for values of $E_1,E_3 \gtrsim
30$~MeV, JUNO loses sensitivity, since the decoherence effects will be
shifted to higher energies and the reactor neutrino spectrum would be
very little modified. We note however, that for such large values of
$E_i$, decoherence effects may show up in long-baseline or atmospheric neutrino
experiments.

Hence, if JUNO does not find any deviation of the energy spectrum from
standard oscillations, our scenario would be highly constrained. A
dedicated investigation of future data from reactor, long-baseline,
and atmospheric neutrino experiments would be required to determine
whether an allowed region survives or not.  In our analysis we have
fixed the exponent in Eq.~\eqref{gam} to $n = 2$. If JUNO would obain
results consistent with standard oscillations, one might also test
values $n > 2$. Increasing $n$, the  decoherence for $E < E_1, E_3$
becomes more strongly suppressed which in turn leads to a faster  weakening of  the sensitivity of JUNO
to the parameter range allowed by LSND. Such investigations are
beyond the scope of the present work. Note also that we have fixed
$\gamma_0 = 0.01$~m$^{-1}$. Smaller values of $\gamma_0$ would not
allow to fit LSND, as discussed in the previous section. For larger
values of $\gamma_0$ decoherence effects will become larger at JUNO,
increasing the sensitivity.

Finally let us mention that the scenario presented here predicts also
large modifications for supernova neutrinos, since the energy range of
supernova neutrinos (tens of MeV) is close to the LSND energy range,
where decoherence effects are important. A discussion of
decoherence effects for supernova neutrinos can be found in
\cite{Farzan:2008zv}.

\section{Conclusions}
\label{sec:conclusion}

We have revisited the idea of quantum decoherence as a solution to the LSND anomaly proposed in \cite{Farzan:2008zv} taking into account the recent results from the Daya Bay and RENO reactor experiments. We assume an exponential dependence of the decoherence parameters on neutrino energy as shown in Eq.~(\ref{di}). For a suitable choice of parameters the decoherence effects can become suppressed for neutrino energies both below and above LSND energies, restricting deviations from the standard three-flavour oscillation scenario to the 20--50~MeV energy range. In this way neither standard oscillations of MeV neutrinos from the Sun and from reactors are modified, nor the results for neutrinos with energies greater than 200~MeV are affected, as relevant for short and long baseline accelerator experiments and atmospheric neutrinos. Moreover the scenario becomes free from the famous appearance--disappearance tension that plagues the 3+1 sterile neutrino solution for the LSND anomaly. We have studied the parameter space in which the LSND anomaly can be explained and constraints from various reactor and short baseline neutrino experiments can be avoided. Results are shown in Fig.~\ref{Ei}.  We have found a remarkable agreement between KARMEN and LSND data within this scenario. The decoherence solution to LSND predicts no effect in MiniBooNE and is hence  consistent with the MiniBooNE null-result for $E_\nu > 475~{\rm MeV}$. However, one should seek  another resolution for the low energy excess observed in MiniBooNE between 200 and 475~MeV as well as for the reactor and Gallium anomalies.

The scenario predicts results consistent with standard three-flavour
oscillations for most of the upcoming long and short-baseline neutrino
oscillation experiments. However, reactor experiments at baselines of
around 50~km such as the JUNO or RENO-50 projects will provide a
crucial test of the scenario for large part of the parameter space.

\subsection*{Acknowledgments}

The authors acknowledge partial support from the  European Union FP7 ITN INVISIBLES (Marie Curie Actions, PITN- GA-2011- 289442). P.B.\ is grateful to the Oskar Klein Centre and the CoPS group at Stockholm University for kind hospitality.



\begin{thebibliography}{99}




\bibitem{Aguilar:2001ty} A.~Aguilar {\it et al.}  [LSND Collaboration],
Phys.\ Rev.\ D {\bf 64}, 112007 (2001) [hep-ex/0104049].


\bibitem{tension}
  J.~Kopp, M.~Maltoni and T.~Schwetz,
  Phys.\ Rev.\ Lett.\  {\bf 107} (2011) 091801
  [arXiv:1103.4570 [hep-ph]];
   J.~Kopp, P.~A.~N.~Machado, M.~Maltoni and T.~Schwetz,
  JHEP {\bf 1305} (2013) 050
  [arXiv:1303.3011 [hep-ph]].

\bibitem{Giunti:2013aea}
  C.~Giunti, M.~Laveder, Y.~F.~Li and H.~W.~Long,
  Phys.\ Rev.\ D {\bf 88} (2013) 073008
  [arXiv:1308.5288 [hep-ph]].

\bibitem{Planck:2015xua}
  P.~A.~R.~Ade {\it et al.}  [Planck Collaboration],
  arXiv:1502.01589 [astro-ph.CO].

\bibitem{cosmol}
  J.~Bergstr\"om, M.~C.~Gonzalez-Garcia, V.~Niro and J.~Salvado,
  JHEP {\bf 1410} (2014) 104
  [arXiv:1407.3806 [hep-ph]];
%
  S.~Gariazzo, C.~Giunti and M.~Laveder,
  JHEP {\bf 1311} (2013) 211
  [arXiv:1309.3192 [hep-ph]].

\bibitem{Gabriela-04}
G.~Barenboim and N.~E.~Mavromatos,
JHEP {\bf 0501} (2005) 034
[hep-ph/0404014].

\bibitem{Gabriela-06}
G.~Barenboim, N.~E.~Mavromatos, S.~Sarkar and A.~Waldron-Lauda,
Nucl.\ Phys.\  B {\bf 758} (2006) 90
[hep-ph/0603028].

\bibitem{Farzan:2008zv}
Y.~Farzan, T.~Schwetz and A.~Y.~Smirnov,
JHEP {\bf 0807} (2008) 067
[arXiv:0805.2098 [hep-ph]].

\bibitem{Giddings}
  J.~R.~Ellis, J.~S.~Hagelin, D.~V.~Nanopoulos and M.~Srednicki,
  Nucl.\ Phys.\  B {\bf 241} (1984) 381;
  S.~B.~Giddings and A.~Strominger,
  Nucl.\ Phys.\  B {\bf 307} (1988) 854;
  N.~E.~Mavromatos and S.~Sarkar,
  hep-ph/0612193.


\bibitem{NuTeV}
S.~Avvakumov {\it et al.},
  Phys.\ Rev.\ Lett.\  {\bf 89} (2002) 011804
  [hep-ex/0203018].


\bibitem{An:2013zwz}
  F.~P.~An {\it et al.}  [Daya Bay Collaboration],
  Phys.\ Rev.\ Lett.\  {\bf 112} (2014) 061801
  [arXiv:1310.6732 [hep-ex]].

\bibitem{An:2013uza}
  F.~P.~An {\it et al.}  [Daya Bay Collaboration],
  Chin.\ Phys.\ C {\bf 37} (2013) 011001
  [arXiv:1210.6327 [hep-ex]].

\bibitem{reno}
  J.~K.~Ahn {\it et al.}  [RENO Collaboration],
  Phys.\ Rev.\ Lett.\  {\bf 108} (2012) 191802
  [arXiv:1204.0626 [hep-ex]];
%
\bibitem{Kim:2014rfa}
  S.~B.~Kim,
  arXiv:1412.2199 [hep-ex].

\bibitem{Abe:2014bwa}
  Y.~Abe {\it et al.}  [Double Chooz Collaboration],
  JHEP {\bf 1410} (2014) 086
   [Erratum-ibid.\  {\bf 1502} (2015) 074]
  [arXiv:1406.7763 [hep-ex]].

\bibitem{Lindblad}
G.~Lindblad,
Commun.\ Math.\ Phys.\  {\bf 48} (1976) 119.

\bibitem{Banks}
T.~Banks, L.~Susskind and M.~E.~Peskin,
Nucl.\ Phys.\  B {\bf 244} (1984) 125.

\bibitem{Maki:1962mu}
Z.~Maki, M.~Nakagawa and S.~Sakata,
Prog.\ Theor.\ Phys.\  {\bf 28} (1962) 870.
%
B.~Pontecorvo,
Zh.\ Eksp.\ Teor.\ Fiz.\ {\bf 53}, 1717 (1967)
[Sov.\ Phys.\ JETP {\bf 26}, 984 (1968)].


\bibitem{Adamson:2014vgd}
  P.~Adamson {\it et al.}  [MINOS Collaboration],
  Phys.\ Rev.\ Lett.\  {\bf 112} (2014) 191801
  [arXiv:1403.0867 [hep-ex]].

\bibitem{Abe:2013hdq}
  K.~Abe {\it et al.}  [T2K Collaboration],
  Phys.\ Rev.\ Lett.\  {\bf 112} (2014) 061802
  [arXiv:1311.4750 [hep-ex]];
%
  Phys.\ Rev.\ Lett.\  {\bf 112} (2014) 18,  181801
  [arXiv:1403.1532 [hep-ex]];
%
  arXiv:1502.01550 [hep-ex].

\bibitem{Ashie:2004mr}
  Y.~Ashie {\it et al.}  [Super-Kamiokande Collaboration],
  Phys.\ Rev.\ Lett.\  {\bf 93} (2004) 101801
  [hep-ex/0404034];
%
  R.~Wendell {\it et al.}  [Super-Kamiokande Collaboration],
  Phys.\ Rev.\ D {\bf 81} (2010) 092004
  [arXiv:1002.3471 [hep-ex]].


\bibitem{AguilarArevalo:2007it}
  A.~A.~Aguilar-Arevalo {\it et al.}  [MiniBooNE Collaboration],
  Phys.\ Rev.\ Lett.\  {\bf 98} (2007) 231801
  [arXiv:0704.1500 [hep-ex]];
%
  Phys.\ Rev.\ Lett.\  {\bf 110} (2013) 161801
  [arXiv:1207.4809 [hep-ex], arXiv:1303.2588 [hep-ex]].

\bibitem{Mahn:2011ea}
  K.~B.~M.~Mahn {\it et al.}  [SciBooNE and MiniBooNE Collaborations],
  Phys.\ Rev.\ D {\bf 85} (2012) 032007
  [arXiv:1106.5685 [hep-ex]].


\bibitem{Dydak:1983zq}
  F.~Dydak {\it et al.},
  Phys.\ Lett.\ B {\bf 134}, 281 (1984).

\bibitem{NOMAD}
  P.~Astier {\it et al.}  [NOMAD Collaboration],
  Phys.\ Lett.\  B {\bf 570} (2003) 19
  [hep-ex/0306037].




\bibitem{kamland}
T.~Araki {\it et al.}, [KamLAND Collaboration],
Phys.\ Rev.\ Lett.\  {\bf 94}, 081801 (2005)
[hep-ex/0406035].

\bibitem{karmen}
B.~Armbruster {\it et al.}  [KARMEN Collaboration],
Phys.\ Rev.\ D {\bf 65}, 112001 (2002)
[hep-ex/0203021].




\bibitem{Gonzalez-Garcia:2014bfa}
  M.~C.~Gonzalez-Garcia, M.~Maltoni and T.~Schwetz,
  JHEP {\bf 1411} (2014) 052
  [arXiv:1409.5439 [hep-ph]].


\bibitem{Maltoni:2002xd}
M.~Maltoni, T.~Schwetz, M.~A.~Tortola and J.~W.~F.~Valle,
Nucl.\ Phys.\ B {\bf 643}, 321 (2002)
[hep-ph/0207157].

\bibitem{Maltoni:2003cu}
M.~Maltoni and T.~Schwetz,
Phys.\ Rev.\ D {\bf 68}, 033020 (2003)
[hep-ph/0304176].

\bibitem{AguilarArevalo:2008rc}
  A.~A.~Aguilar-Arevalo {\it et al.}  [MiniBooNE Collaboration],
  Phys.\ Rev.\ Lett.\  {\bf 102} (2009) 101802
  [arXiv:0812.2243 [hep-ex]].



\bibitem{rec-an}
T.~.A.~Mueller {\it et al.}, 
 Phys.\ Rev.\ C {\bf 83} (2011) 054615  [arXiv:1101.2663 [hep-ex]];
%
  P.~Huber,
  Phys.\ Rev.\ C {\bf 84} (2011) 024617
   [Erratum-ibid.\ C {\bf 85} (2012) 029901]
  [arXiv:1106.0687 [hep-ph]];
%
  G.~Mention {\it et al.}, 
  Phys.\ Rev.\ D {\bf 83} (2011) 073006
  [arXiv:1101.2755 [hep-ex]].


\bibitem{Giunti:2010zu}
  C.~Giunti and M.~Laveder,
  Phys.\ Rev.\ C {\bf 83} (2011) 065504
  [arXiv:1006.3244 [hep-ph]].



\bibitem{Patterson:2012zs}
  R.~B.~Patterson [NOvA Collaboration],
  Nucl.\ Phys.\ Proc.\ Suppl.\  {\bf 235-236} (2013) 151
  [arXiv:1209.0716 [hep-ex]].

\bibitem{lbnf}
  C.~Adams {\it et al.}  [LBNE Collaboration],
  arXiv:1307.7335 [hep-ex]; The Long Baseline Neutrino Facillity (LBNF),
  {\tt https://web.fnal.gov/project/LBNF/SitePages/Home.aspx}

\bibitem{Adey:2014rfv}
  D.~Adey {\it et al.}  [nuSTORM Collaboration],
  Phys.\ Rev.\ D {\bf 89} (2014) 7,  071301
  [arXiv:1402.5250 [hep-ex]].

\bibitem{sbl-future}
R.~Acciarri {\it et al.}  [MicroBooNE and LAr1-ND and ICARUS-WA104 Collaborations],
  arXiv:1503.01520 [physics.ins-det];
  A.~Anokhina {\it et al.}, 
  arXiv:1404.2521 [hep-ph].

\bibitem{ess}
  E.~Baussan {\it et al.}  [ESSnuSB Collaboration],
  Nucl.\ Phys.\ B {\bf 885} (2014) 127
  [arXiv:1309.7022 [hep-ex]];
  M.~Blennow, P.~Coloma and E.~Fernandez-Martinez,
  JHEP {\bf 1412} (2014) 120
  [arXiv:1407.1317 [hep-ph]].


\bibitem{Elnimr:2013wfa}
  M.~Elnimr {\it et al.}  [OscSNS Collaboration],
  arXiv:1307.7097.


\bibitem{He:2014zwa}
  M.~He [JUNO Collaboration],
  arXiv:1412.4195 [physics.ins-det].

\bibitem{Li:2013zyd}
  Y.~F.~Li, J.~Cao, Y.~Wang and L.~Zhan,
  Phys.\ Rev.\ D {\bf 88} (2013) 013008
  [arXiv:1303.6733 [hep-ex]].


\bibitem{Blennow:2013oma}
  M.~Blennow, P.~Coloma, P.~Huber and T.~Schwetz,
  JHEP {\bf 1403} (2014) 028
  [arXiv:1311.1822 [hep-ph]].

\end{thebibliography}
\end{document}